\documentclass[runningheads]{llncs}
\usepackage[T1]{fontenc}
\usepackage{graphicx}
\usepackage{hyperref}
\usepackage{url}
\usepackage{stackengine}

\usepackage{enumerate}
\usepackage{enumitem}
\newcommand{\Diff}[0]{\mathrm{Diff}}    

\usepackage{booktabs}
\usepackage{multirow} 
\usepackage{amsmath}
\usepackage{ulem}
\usepackage{xcolor}
\usepackage{amssymb}
\usepackage{amsfonts}
\usepackage{lmodern}
\usepackage{mathrsfs}
\usepackage{bm}
\usepackage{booktabs}    
\usepackage{diagbox} 
\usepackage{makecell} 
\usepackage{relsize}
\usepackage{algorithm} 
\usepackage{algorithmic}
\usepackage{wrapfig}
\begin{document}
%
\title{IGG: Image Generation Informed by Geodesic Dynamics in Deformation Spaces}
\author{Nian Wu  \inst{1} 
\and
Nivetha Jayakumar \inst{1} 
\and
Jiarui Xing\inst{1} 
\and
Miaomiao Zhang\inst{1,2}} 
\authorrunning{Wu and Zhang}
\institute{
Department of Electrical and Computer Engineering, University of Virginia, USA \and
Department of Computer Science, University of Virginia, USA}

\maketitle              
\begin{abstract}
Generative models have recently gained increasing attention in image generation and editing tasks. However, they often lack a direct connection to object geometry, which is crucial in sensitive domains such as computational anatomy, biology, and robotics. This paper presents a novel framework for Image Generation informed by Geodesic dynamics (IGG) in deformation spaces. Our IGG model comprises two key components: (i) an efficient autoencoder that explicitly learns the geodesic path of image transformations in the latent space; and (ii) a latent geodesic diffusion model that captures the distribution of latent representations of geodesic deformations conditioned on text instructions. By leveraging geodesic paths, our method ensures smooth, topology-preserving, and interpretable deformations, capturing complex variations in image structures while maintaining geometric consistency. We validate the proposed IGG on plant growth data and brain magnetic resonance imaging (MRI). Experimental results show that IGG outperforms the state-of-the-art image generation/editing models with superior performance in  generating realistic, high-quality images with preserved object topology and reduced artifacts. Our code is publicly available at https://github.com/nellie689/IGG.
\end{abstract}
\section{Introduction}
Generative diffusion models have gained significant attention in recent years due to their ability to capture complex data distributions and produce high-quality and realistic samples~\cite{brock2019large,mukhopadhyay2023diffusion}. Among these, language-image generative models \cite{ramesh2022hierarchical,saharia2022photorealistic} leverage text prompts to guide the diffusion process, allowing users to perform highly customizable and context-aware edits to images. In medical imaging, such models are widely utilized to create diverse datasets, improving downstream tasks such as image classification~\cite{yang2023diffmic,ijishakin2023interpretable}, segmentation~\cite{amit2021segdiff,fernandez2022can}, and object recognition~\cite{hamamci2023diffusion}. Additionally, they achieve great performance in image-to-image translation, addressing key challenges in the synthesis of high-quality medical scans from low-quality data~\cite{chung2022score,mao2023disc} and the simulation of missing modalities ~\cite{zhu2023make,ozbey2023unsupervised}. With their capability to potentially improve disease diagnostic accuracy and support personalized treatment planning, generative diffusion models hold great promise for advancing real-world clinical applications in healthcare.

Despite their success, existing diffusion models primarily focus on manipulating the intensity and texture features extracted from images, often neglecting direct connections to object geometry in the editing or generation process~\cite{nguyen2024visual,li2024blip,hertz2022prompttoprompt,kawar2023imagic}. This limitation can lead to unrealistic or biologically implausible samples, posing risks to downstream tasks that require accurate geometric preservation~\cite{azizi2023synthetic}. Recent advances have introduced simple geometric constraints such as shape localization~\cite{patashnik2023localizing,gupta2024topodiffusionnet}, boundary conditions~\cite{maze2023diffusion}, and 3D shape priors~\cite{yu2025surf,hu2024topology} in the modeling process. However, these approaches lack fine-grained structural preservation, which is highly desirable in sensitive domains, including computational anatomy, robotics, and medical imaging. Another important yet under-investigated limitation of current diffusion models is their inability to provide interpretable metrics or quantitative measures of topological integrity for image objects. Commonly used evaluation metrics include Fréchet Inception Distance (FID) for assessing distribution similarity between real and generated data~\cite{heusel2017gans}, Inception Score (IS) for measuring diversity and quality~\cite{salimans2016improved}, and Structural Similarity Index (SSIM) for evaluating pixel-level fidelity~\cite{wang2004image}. While these metrics effectively assess visual quality, they do not account for the preservation of object geometry and topology, raising questions about their reliability and trustworthiness in structure-sensitive applications.

To overcome these limitations, we introduce a novel framework, Image Generation informed by Geodesic dynamics (IGG), that for the first time generates images by deforming a given template along random geodesics in deformation spaces guided by text instructions. In contrast to current approaches~\cite{brooks2023instructpix2pix,ho2022video,kim2022diffusemorph}, our IGG ensures topological consistency by treating each generated image as a deformed variation of a template or reference image through learned diffeomorphic transformations (i.e., a one-to-one, smooth, and invertible mapping). A dynamic geodesic, producing natural transformations and providing interpretable metrics to quantify topological changes, will be explicitly learned during the diffusion process. Our proposed IGG model comprises two key components: a latent representation learning framework of diffeomorphic transformations informed by geodesic dynamics, and a latent geometric diffusion model conditioned on user-defined text instructions. To summarize, the contributions of IGG are threefold:
\begin{enumerate}[label=(\roman*)]
\item Develop a novel image generation model that integrates geometric principles through the geodesic learning of image deformations with a powerful conditional diffusion process.
\item Enable quantitative metrics to assess and ensure the topology consistency of generated samples.
\item Establish a new paradigm for producing anatomically intact and realistic image samples, potentially advancing applications in domains requiring structural fidelity and precision.
\end{enumerate}

We demonstrate the effectiveness of IGG on a diverse set of real-world datasets, including Komatsuna plant growth data~\cite{uchiyama2017easy} and brain MRIs~\cite{lamontagne2019oasis}. We compare the performance of the model with state-of-the-art text-instructed generative models. Experimental results show that IGG achieves significantly improved results in generating image samples with well-preserved topological structures.

\section{Background: Geodesics In Deformation Spaces}
\label{sec:backgroundlddmm}
In this section, we briefly review the basic concept of geodesics in deformation spaces, which provides a principled way to model smooth and natural deformations between images while preserving structural integrity~\cite{miller2002metrics}. With the underlying assumption that objects of a generic class (e.g., human brains, hearts, or lungs) are described as deformed variants of a given template, descriptors of that class arise naturally by deforming the template to other images along a {\bf geodesic} - a shortest path with minimal energy of transformations~\cite{avants2008symmetric,joshi2004unbiased}. 
In theory, every topological property of the deformed template can be preserved by enforcing the resulting transformations to be diffeomorphisms, i.e., differentiable, bijective mappings with differentiable inverses~\cite{beg2005computing,arnold1966geometrie,miller2002metrics}. Violations of such constraints introduce image artifacts, such as tearing, crossing, or passing through itself. 

Let $\Diff^\infty(\Omega)$ denote the space of smooth
diffeomorphisms on a $d$-dimensional torus domain $\Omega = \mathbb{R}^d / \mathbb{Z}^d$. The tangent space of diffeomorphisms is the
space $V = \mathfrak{X}^\infty(T\Omega)$ of smooth vector fields on $\Omega$. Considering a time-varying velocity field, $\{v_t\} : [0,\tau] \rightarrow V$, we can generate diffeomorphisms $\{\phi_{t}\}$ between pairwise images by solving
\begin{equation}
\label{eq:phi_v}
\frac{d \phi_t}{dt} = v_t(\phi_t), \, t \in [0, 1], 
\end{equation}
where $\phi_0$ is the identity map and $\phi_1$ is the target transformation.

The geodesic minimizes the functional $\int_0^1 (\mathcal{L} v_t, v_t) \, dt$~\cite{beg2005computing}, where $\mathcal{L}: V\rightarrow V^{*}$ is a symmetric, positive-definite differential operator that maps a tangent vector $ v(t)\in V$ into its dual space as a momentum vector $m(t) \in V^*$. We typically write $m(t) = \mathcal{L} v(t)$, or $v(t) = \mathcal{K} m(t)$, with $\mathcal{K}$ being an inverse operator of $\mathcal{L}$. In this paper, we adopt a commonly used Laplacian operator $\mathcal{L}=(- \alpha \Delta + \text{Id})^3$, where $\alpha$ is a weighting parameter that controls the smoothness of transformation fields and $\text{Id}$ is an identity matrix. The $(\cdot, \cdot)$ is a dual paring, which is similar to an inner product between vectors. According to a well-known geodesic shooting algorithm~\cite{vialard2012}, the minimum of the functional mentioned above is uniquely determined by solving a Euler-Poincar\'{e} differential (EPDiff) equation~\cite{vialard2012,younes2009evolutions} with a given initial condition. That is, for $\forall v_0 \in V$, a geodesic path $t \mapsto \phi_t$ in the space of diffeomorphisms can be computed by forward shooting the EPDiff equation
\begin{equation}
\label{eq:epdiff}
    \frac{\partial v_t}{\partial t} =-K\left[(Dv_t)^Tm_t + Dm_t\, v_t + m_t \operatorname{div} v_t\right],
\end{equation}
where $D$ denotes the Jacobian matrix and $\operatorname{div}$ is a divergence operator.

\noindent {\bf Derive geodesics from images.} Consider a source image $S$ and a target image $T$ defined in the domain $\Omega$ ($S(x), T(x):\Omega \rightarrow \mathbb{R}$). The optimization of geodesic transformations can be formulated as minimizing an energy function over the initial velocity, subject to the EPDiff equations, i.e., 
\begin{equation*}
\label{eq:lddmm}
 E(v_0) = (\mathcal{L} v_0, v_0)  + \lambda  \text{Dist}(S(\phi_1), T)  \, \, \text{s.t. Eq.}~\eqref{eq:phi_v}  \, \& \, ~\eqref{eq:epdiff}.
\end{equation*}
Here, Dist(·,·) is a distance function that measures the dissimilarity between images and $\lambda$ is a positive weighting parameter. In this paper, we will use the commonly used sum-of-squared intensity differences~\cite{beg2005computing,wu2024tlrn,zhang2017frequency}. 

\section{Our Method: IGG}
\begin{figure*}[!b]
\centering
\includegraphics[width=\textwidth]{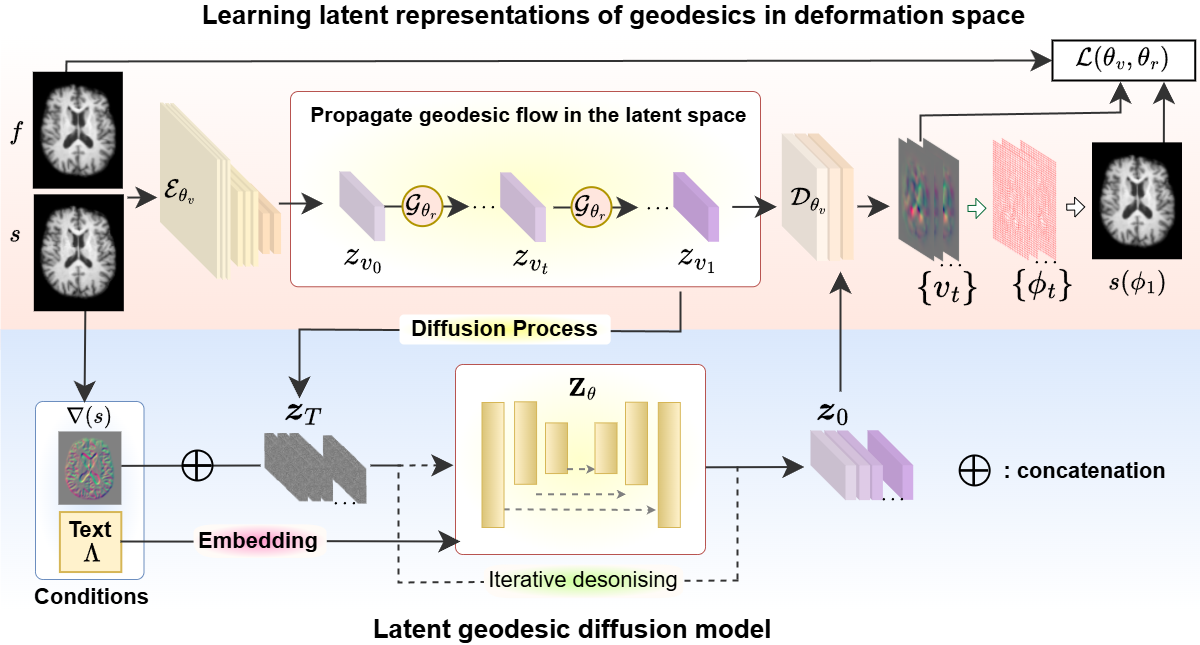}
\caption{An overview of the proposed IGG model.}
\label{fig:Arc}
\end{figure*}

This section introduces our proposed model, IGG, that for the first time generates images in deformation spaces guided by text instructions. A dynamic geodesic, producing natural transformations and providing interpretable metrics to quantify topological changes, will be explicitly learned during the diffusion process. Our proposed IGG model consists of two main components: (i) an autoencoder-based deformable registration network that learns the latent representations of diffeomorphic transformations informed by geodesic dynamics; and (ii) a latent geodesic diffusion model that captures the latent distribution of sequential deformation features, conditioned on user-defined text inputs. An overview of our model is illustrated in Fig.~\ref{fig:Arc}. 

\subsection{Network Architecture}
\paragraph*{\bf Learning latent representations of geodesics in deformation space.} For a number of $N$ template and target images associated with text instructions $\{s^n, f^n, \Lambda^n\}_{n=1}^N$, we employ an autoencoder-based deformable registration network to learn the latent representation of geodesic transformations, parameterized by velocity fields $\{v_t\}$, between a template $s^n$ and a target $f^n$. The encoder of this network, $\mathcal{E}_{\theta_v}$, maps input data to the latent representation of a sequence of velocity fields, $\{z_{v_t}\}$. Inspired by the recent work NeurEPDiff~\cite{wu2023neurepdiff}, which develops a neural operator for learning the mapping functions of the EPDiff equation, we incorporate the NeurEPDiff module, $\mathcal{G}_{\theta_r}$, in our model IGG to learn geodesics in latent deformation spaces. More specifically, starting with a latent initial velocity $z_{v_0}$, the operator $\mathcal{G}_{\theta_r}$ iteratively propagates it along the geodesic path, $z_{v_0} \longmapsto \cdots z_{v_t} \longmapsto \cdots z_{v_1}$. At each time point $t$, a multilayer neural network is employed to simulate the geodesic mapping function from $z_{v_{t}}$ to $z_{v_{t+1}}$. Each hidden layer combines a local linear transformation, $W^j$, and a global convolutional kernel, $\mathcal{H}^j$, to extract both global and local representations. Additionally, a smooth nonlinear activation function, $\sigma(\cdot)$, is used to encourage a smooth evolution of the geodesic path. This process is defined as
\begin{equation*}
z_{v_{t+1}}:= \sigma(W^J, \mathcal{H}^J) \circ \cdots \circ \sigma(W^2, \mathcal{H}^2) \circ \sigma(W^1, \mathcal{H}^1, z_{v_t}),
\label{eq:FneurEPDiff}
\end{equation*}
where $\circ$ denotes the composition of network operations. The decoder, $\mathcal{D}_{\theta_v}$, then projects these latent representations back to the full-dimensional image space to produce a geodesic flow of velocity fields, $v_0 \longmapsto \cdots v_t \longmapsto \cdots v_1$, which are integrated to generate the corresponding deformations, $\phi_0 \longmapsto \cdots \phi_t \longmapsto \cdots \phi_1$ as described in Eq.~\ref{eq:phi_v}.

The network loss combines contributions from an unsupervised registration loss, and a geodesic loss guided by numerical solutions of the EPDiff equation obtained via Euler integration. These solutions are represented as $\{\hat{v}_t^n \}, t \in (0, 1]$ with a simultaneously learned initial velocity $\hat{v}_{0}^n$. To simplify the notation, we define $\bm{\hat{v}}^n \triangleq \{ \hat{v}_t^n \}$ before formulating the loss function as
\begin{align}
\label{eq:JointLossFunGeodesic}
\mathcal{L}(\theta_v, \theta_r) &=  \sum_{n=1}^{N} \lambda \, \|(s^n(\phi^{n}_1 (\theta_v)) - f^n \|^2_2 
+ \frac{1}{2} (\mathcal{L} v^n_0 (\theta_v), v^n_0 (\theta_v))  \nonumber \\
&+ \eta \, \| \mathcal{D}_{\theta_v} (\mathcal{G}_{\theta_r}(z_{v_0}))- \bm{\hat{v}}^n \|_2^2 + \text{Reg}(\theta_r, \theta_v),
\,\,\,\, s.t. \,\, \text{Eq.} ~\eqref{eq:phi_v}, 
\end{align}
where $\lambda$ and $\eta$ are positive weighting parameters to balance the image matching term and the geodesic loss, and $\text{Reg}(\cdot)$ is a network regularization. 

\paragraph*{\bf Latent geodesic diffusion model.} With a pre-trained geodesic learning network, we are now ready to introduce a latent geometric diffusion module to simulate the distribution of latent geodesic flows for image transformations, \(\{ z^n_{v_t}\} \). To simplify the notation, we define $\boldsymbol{z}^{n} := \{ z^n_{v_t}\}_{t=0}^{1}$ for the $n$-th subject, where all time-dependent components $\{z^n_{v_t}\}$ are concatenated to form a unified representation. In contrast to previous approaches that perform the diffusion process in the image space~\cite{brooks2023instructpix2pix,ho2022video,kim2022diffusemorph}, our latent geometric diffusion operates in the deformation space. Specifically, our diffusion process directly samples geodesics of topology-preserving diffeomorphic transformations, which are then applied to deform the input template to generate the final output. \\

\noindent \textit{Geometric conditioning in the latent geodesic space.} Following the principles of video diffusion models~\cite{ho2022video}, we perform forward diffusion process by progressively adding noise, $\epsilon \sim \mathcal{N}(0, \mathbf{I})$, to the initial latent representations $\boldsymbol{z}^{n}_0 \sim p(\boldsymbol{z}^{n}_0), \boldsymbol{z}^{n}_{0} \leftarrow \boldsymbol{z}^{n}$ with a number of $T$ steps. The forward diffusion process results in a distribution $q(\boldsymbol{z}^{n}_{\tau} \vert \boldsymbol{z}^{n}_0)$, where $\tau \in [1, \cdots, T]$~\cite{ho2022video}.

In the reverse diffusion process, we first introduce a novel geometric conditioning in the latent geodesic space. This enables our model to adapt more effectively to specific contexts; hence improving its predictive performance based on observed template images and provided text instructions. More specifically, our geometric conditioning is achieved by concatenating three components: image embedding - downsampled template image gradient, $\{\nabla s^n\}$; the text embedding, encoded using a CLIP text encoder~\cite{radford2021learning}, $\{\Lambda^n\}$; and the geodesic deformation embedding, represented by the latent velocity fields, $\{\boldsymbol{z}_{\tau}^n\}$. The reverse diffusion process then involves sampling the denoised latent geodesic flow of velocity fields $\hat{\boldsymbol{z}}^n_0$, starting from the noisy state $\boldsymbol{z}^n_{\tau}$ at the $\tau$-th time-step and progressively removing noise to reconstruct the original latent representation. This process is mathematically expressed as
\begin{equation}
    p_\theta(\boldsymbol{z}^{n}_{\tau-1} | \boldsymbol{z}^{n}_{\tau}, \nabla s^{n} , \Lambda^{n}, \tau) = \mathcal{N}(\boldsymbol{z}^{n}_{\tau-1}; \mu_\theta(\boldsymbol{z}^{n}_{\tau}, \nabla s^{n} , \Lambda^{n}, \tau),\Sigma_\theta(\boldsymbol{z}^{n}_{\tau}, \nabla s^n , \Lambda^n, \tau)),
\end{equation}
where $\mu_\theta$ \cite{ho2022video} and $\Sigma_\theta$ represent the mean and variance of the Gaussian distribution at each time step of the reverse diffusion process. The parameters $\mu_\theta$ \cite{ho2022video} and $\Sigma_\theta$ are learned functions, and the reverse process is practically implemented using a denoising network, $\mathbf{Z}_\theta$. In our experiments, we employ a 3D UNet architecture~\cite{ronneberger2015u} as the backbone for ${\mathbf{Z}_\theta}$, which predicts the noise component to be removed from $\boldsymbol{z}^{n}_{\tau}$ to reconstruct $\boldsymbol{z}^{n}_{\tau-1}$.

The network loss of our proposed latent geodesic diffusion module is formulated as
\begin{equation}
\label{eq:IGG}
\mathcal{L}_{\theta} = \frac{1}{N}\mathlarger{\sum}_{n=1}^N \vert\vert \epsilon_{\tau}^n - {\mathbf{Z}}_{\theta}(\boldsymbol{z}^{n}_{\tau}, \nabla s^n , \Lambda^n, \tau) \vert\vert^2_2  + \text{reg}(\theta), \tau \in \text{U}[1,T],
\end{equation}
where $\text{reg}(\cdot)$ is a regularization of the network parameter $\theta$, and U denotes a uniform  distribution.

To balance the trade-off between sample quality and diversity of velocity fields in the latent space after training, we optimize the combination of conditional and unconditional diffusion models. This is achieved by leveraging image-conditioned guidance with scale $\delta_\mathrm{I}$, text-conditioned guidance with scale $\delta_\mathrm{T}$, and a null-condition $\varnothing$. Following similar principles from the classifier-free guidance approach~\cite{ho2022classifier}, we first jointly train a conditional and an unconditional IGG model. The resulting score estimates from these models are then combined to predict the noise $\hat{\mathbf{Z}}$ as
\begin{align}
\label{eq:cfg}  
\hat{{\mathbf{Z}}}_{\theta}(\boldsymbol{z}^{n}_{\tau}, \nabla s^n , \Lambda^n, \tau) &= {\mathbf{Z}}_{\theta}(\boldsymbol{z}^{n}_{\tau}, \varnothing, \varnothing, \tau) + 
 \delta_\mathrm{I} \cdot \left[{\mathbf{Z}}_{\theta}(\boldsymbol{z}^{n}_{\tau}, \nabla s^n , \varnothing, \tau) -  {\mathbf{Z}}_{\theta}(\boldsymbol{z}^{n}_{\tau}, \varnothing, \varnothing, \tau)\right] \nonumber \\ 
 &\quad + \delta_\mathrm{T} \cdot \left[{\mathbf{Z}}_{\theta}(\boldsymbol{z}^{n}_{\tau}, \nabla s^n , \Lambda^n, \tau) - {\mathbf{Z}}_{\theta}(\boldsymbol{z}^{n}_{\tau}, \nabla s^n, \varnothing, \tau)\right] \nonumber \\
 &= (1-\delta_\mathrm{I}) \cdot {\mathbf{Z}}_{\theta}(\boldsymbol{z}^{n}_{\tau}, \varnothing, \varnothing, \tau) + (\delta_\mathrm{I} - \delta_\mathrm{T}) \cdot {\mathbf{Z}}_{\theta}(\boldsymbol{z}^{n}_{\tau}, \nabla s^n , \varnothing, \tau) \nonumber \\ 
 &\quad + \delta_\mathrm{T} \cdot {\mathbf{Z}}_{\theta}(\boldsymbol{z}^{n}_{\tau}, \nabla s^n , \Lambda^n, \tau).
\end{align}

\subsection{Network Optimization}  
We first optimize the loss of representation learning of geodesic transformations (Eq.~\eqref{eq:JointLossFunGeodesic}), followed by the latent geodesic diffusion model (Eq.~\eqref{eq:IGG}). The training process of IGG is summarized in Alg.~\ref{alg:training}. During testing, the procedure for sampling the geodesic of latent velocity fields, conditioned on a given template image and text instructions, is outlined in Alg.~\ref{alg:sampling}.
\begin{figure}[!h]
\begin{minipage}[t]{0.5\textwidth}
\vspace{-2em}
    \begin{algorithm}[H] 
        \caption{IGG Training}
        \label{alg:training}
        \begin{algorithmic}[1]
        \STATE \textbf{Input: } data $\{ s^n, f^n, \Lambda^n \}_{n=1}^{N_{train}}$
        \STATE Pretrain $\mathcal{E}_{\theta_v}, \mathcal{E}_{\theta_r}, \mathcal{D}_{\theta_v}$ via Eq.~\ref{eq:JointLossFunGeodesic}.
        \REPEAT
        \STATE $\boldsymbol{z}^{n}_0 \gets \mathcal{G}_{\theta_r}(\mathcal{E}_{\theta_v}$($s^n, f^n))$
        \STATE Sample $\tau \sim \text{U}(1, \mathrm{T})$, $\epsilon \sim \mathcal{N}(0, \textbf{I})$
        \STATE $\boldsymbol{z}^{n}_{\tau} \gets q(\boldsymbol{z}^{n}_{\tau} \vert \boldsymbol{z}^{n}_0)$
        \STATE Take gradient descent step on Eq.~\ref{eq:IGG} \\
        \UNTIL{converge}
        \end{algorithmic}
    \end{algorithm}
\end{minipage}
\hfill
\begin{minipage}[t]{0.5\textwidth}
\vspace{-2em}
    \begin{algorithm}[H] 
        \caption{IGG Sampling}
        \label{alg:sampling}
        \begin{algorithmic}[1]
        \STATE \textbf{Input:} data $\{s^n, \Lambda^n\}_{n=1}^{N_{test}}$
        \STATE Initialize $\boldsymbol{z}^{n}_{T} \sim \mathcal{N}(0,\textbf{I})$
        \FOR{$\tau = \mathrm{T},...,1$}
        \STATE $\hat{\boldsymbol{z}}^{n}_{\tau-1} \gets p_{\theta}(\boldsymbol{z}_{\tau-1}^{n}  \vert \boldsymbol{z}^{n}_{\tau}, \nabla s^{n} , \Lambda^{n}, \tau)$
        \ENDFOR
        \STATE $\{\hat{\boldsymbol{v}}^n\} = \mathcal{D}_{{\theta}_{v}}(\hat{\boldsymbol{z}}^{n}_{0})$
        \STATE $\{\phi^n\} \gets \{\hat{\boldsymbol{v}}^n\}$ via Eq.~\ref{eq:phi_v}
        \STATE {\bfseries return} $\{\phi^n\}$ 
        \end{algorithmic}
    \end{algorithm}
\end{minipage}
\end{figure}

\section{Experimental Evaluation}
We evaluate the proposed IGG framework using a diverse set of real-world image datasets that capture deformable shape changes over time. We first assess the quality of the learned latent representations of geodesics by comparing them to numerical solutions of the EPDiff equation obtained via Euler integration. 

Next, we evaluate the quality of the generated images deformed by the sampled geodesics guided by given text instructions. These results are compared with state-of-the-art generative models that synthesize image sequences/videos with publicly available training code or fine-tuning options, including the video diffusion model (VDM)~\cite{ho2022video}, CogVideoX~\cite{yang2024cogvideox}, and DynamiCrafter~\cite{xing2025dynamicrafter}. Note that all baselines are trained using sequences of deformed images generated along the geodesic transformations derived from the numerical solutions of EPDiff equation. This approach ensures that the results are independent of the learned geodesics predicted by the first module of the IGG framework, providing a consistent and unbiased evaluation.

\subsection{Dataset}
\noindent{\bf Komatsuna plant.} We include $300$ frames of RGB-D label-maps representing five Komatsuna plants from the publicly available data repository~\cite{uchiyama2017easy}. The label maps cover different leaves that emerge from the bud and grow in size over time, where the plant growth was monitored between $228$-$236$ hours. All data frames were resampled to to $128^2$ and pre-aligned with affine transformations. \\

\noindent{\bf Longitudinal brain MRI.} We include a total of $2618$ T1-weighted longitudinal (time-series) brain MRIs sourced from the Open Access Series of Imaging Studies (OASIS-3) dataset~\cite{lamontagne2019oasis}. This experiment aims to validate our method using longitudinal data that includes scans at varying time intervals for individuals spanning both healthy subjects and Alzheimer's diseases (AD), aged $60$-$90$. Given the scenario that many existing image generation/editing methods with text instructions focus on 2D natural images~\cite{brooks2023instructpix2pix,meng2021sdedit}, we specifically utilize 2D scans derived from this 3D brain data for comparison with state-of-the-art baselines. All MRIs undergo pre-processing, including resizing to $128^2$, with isotropic voxels of $1 \text{mm}^2$, skull-stripping, intensity normalization, bias field correction, and pre-alignment using affine transformations. \\

\noindent{\bf Text instructions.} Our text condition consists of a description of the template image, followed by details of the specific edits or progressions applied to it. For brain MRIs, this includes biological variables such as age, sex, and gender. In contrast, for plants, it primarily focuses on the growth timeline. All information is sourced directly from the original dataset available in the data repository.

\subsection{Experimental Design and Implementation Details}
\noindent{\bf Evaluate learned latent representations of geodesics.} We evaluate the model's ability to predict geodesic dynamics by comparing the learned geodesics with numerical solutions to the EPDiff equation using Euler integration. The assessment focuses on errors in velocity fields, transformations, and deformed images between the two approaches. Quantitative metrics, such as the mean absolute error of velocity fields, are computed at each time integration step. \\

\noindent{\bf Evaluate generated samples.} We compare the quality of samples generated by IGG with three baseline models: VDM~\cite{ho2022video}, CogVideoX~\cite{yang2024cogvideox}, and DynamiCrafter~\cite{xing2025dynamicrafter}). In particular, we first assess individual synthesized images at each time step using standard image synthesis metrics, including the Fréchet Inception Distance (FID)~\cite{heusel2017gans} and Kernel Inception Distance (KID)~\cite{binkowski2018demystifying} to measure the distributional similarity between generated and real images. Additionally, we use the Structural Similarity Index (SSIM)~\cite{wang2004image} to evaluate structural, contrast, and luminance similarities. We then evaluate the continuity and smoothness of generated video sequences by computing the Fréchet Video Distance (FVD)~\cite{unterthiner2018towards} and providing qualitative visual comparisons. Finally, we assess the perceptual quality of the generated samples using the Inception Score (IS)~\cite{salimans2016improved}. \\

\noindent{\bf Evaluate the preservation of object geometry and topology.} To demonstrate the efficiency of our IGG model in preserving fine-grained geometric structure and topology, we first generate image sequences by deforming a template image using the sampled geodesic flow of deformations produced by IGG and compare these results with all baseline models. Secondly, we present the corresponding determinant of the Jacobian (DetJac) maps for the deformations. The DetJac values reveal important patterns of volume change: a value of $1$ indicates no volume change, DetJac $<1$ reflects volume shrinkage, and DetJac $>1$ implies volume expansion. A DetJac value below zero indicates artifacts or singularities in the transformation field, highlighting a failure to maintain the topological integrity. Note that only our IGG model enables such a metric to quantify the topological changes in the generated samples. \\

\noindent{\bf Evaluate reliability of model predictions.} We evaluate the reliability and confidence of IGG in learning the geodesic deformations of geometric shapes over time by computing pixel-wise mean and standard deviation for individual sampled image sequences along the geodesic path. We visualize confidence intervals to highlight regions with 95\% certainty in growth patterns. These intervals are defined by the mean of $1000$ generated samples with pixel-wise bounds that are two standard deviations above and below the mean. \\

\noindent{\bf Parameter Setting.} We split all datasets into $80\%$, $10\%$, and $10\%$ for training, validation, and testing. All experiments are conducted on NVIDIA A100 GPUs. For training the autoencoder for geodesic learning, we alternatively update the encoder-decoder and the latent geodesic neural operator, and finally jointly train both components for $2000$ epochs per stage. We set the weight decay as $1e^{-4}$, batch size as $64$, and a learning rate as $5e^{-4}$. For training the latent geodesic diffusion model, we set the learning rate as $1e^{-4}$, the number of epochs as $6000$, batch size as $36$, and diffusion steps as $500$. All models are trained using the Adam optimizer~\cite{kingma2014adam}.

\subsection{Experimental Results}
\begin{figure*}[!h]
\centering
\includegraphics[width=1.0\textwidth] {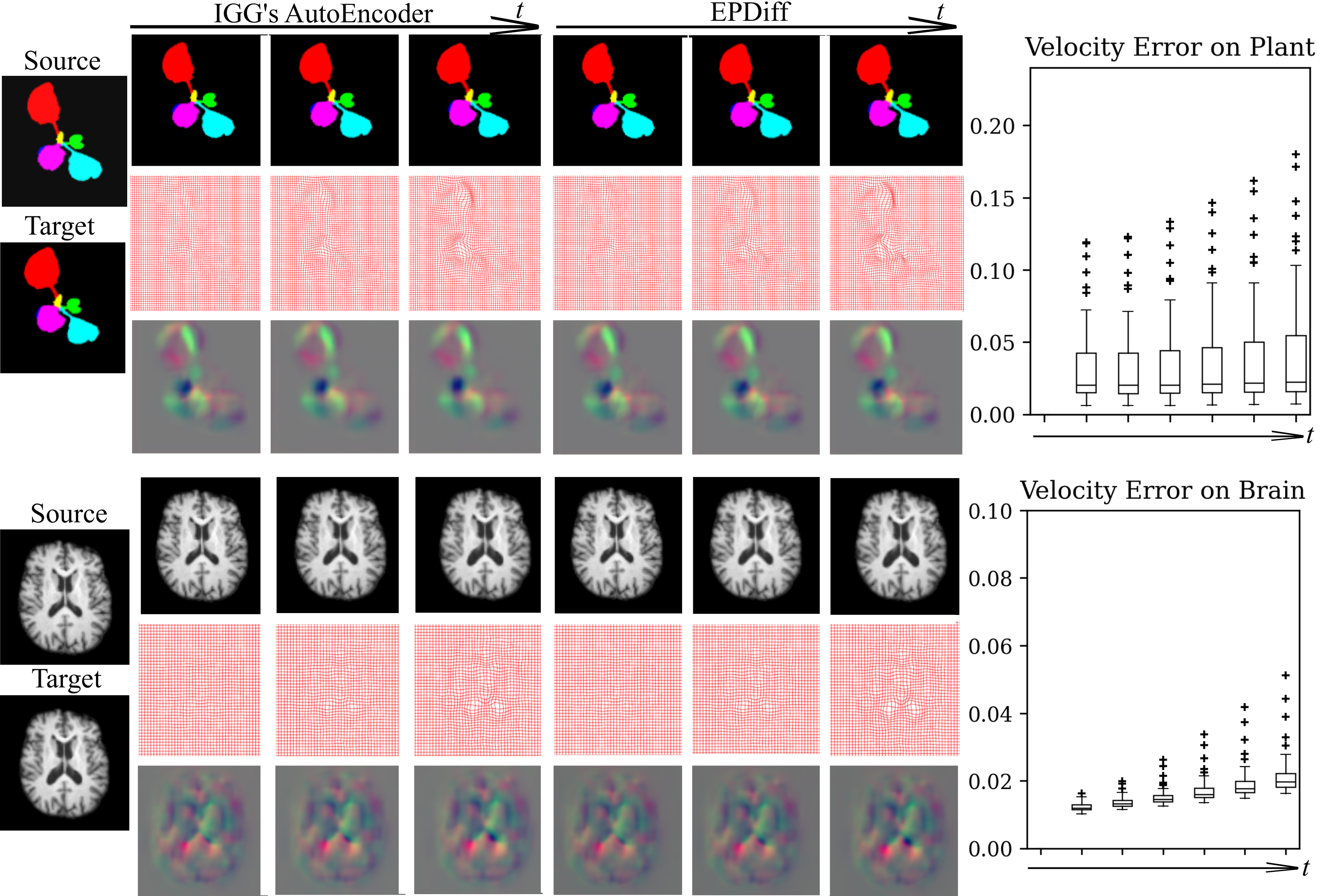}
     \caption{Comparison of predicted geodesics by IGG vs. real numerical solutions from the EPDiff equation. Left: Visualization of predicted deformed images, deformations, and velocities along time. Right: Mean absolute error of predicted velocities over time compared to numerical integration of EPDiff.}
\label{fig:CompareAE}
\end{figure*}

The left panel of Fig.~\ref{fig:CompareAE} visualizes the predicted transformations, velocity fields, and deformed images from IGG's autoencoder, compared with numerical solutions to the EPDiff equation. These methods show a high degree of similarity. The right panel of Fig.~\ref{fig:CompareAE} presents quantitative results of the differences between the real numerical solution and IGG’s latent representation learning module at each time step. These results collectively highlight IGG's ability to effectively learn geodesic mapping functions comparable to real numerical solutions.

\begin{figure*}[!hb]
\centering
\includegraphics[width=0.97\textwidth] {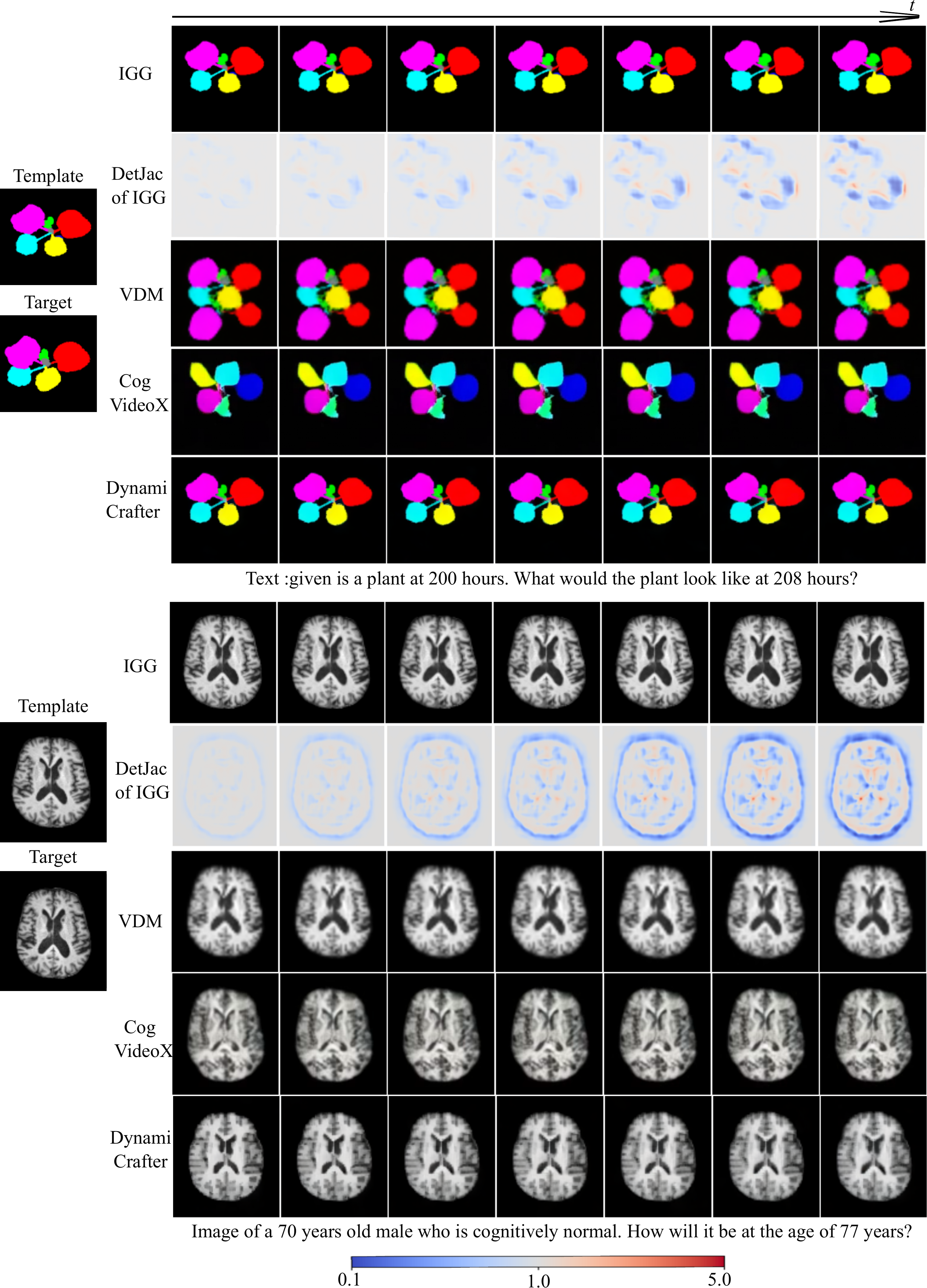}
     \caption{A comparison of images generated by IGG (with corresponding DetJac) against all baseline models across different time frames. Given an input template image and text instructions, all models generate samples of target images. The ground truth "target" along with input template images are provided on the left side of the panel for reference.}
\label{fig:CompareGeo}
\end{figure*}

Fig.~\ref{fig:CompareGeo} compares the generated images from our model IGG (along with the associated DetJac maps of the deformations) with the baseline generative models. The visualized images and DetJac values of IGG suggest that our model effectively preserves the topological structure of objects within the generated images with well-captured progression of geometric shape changes over time. In contrast, samples generated by the baselines fail to maintain the geometric integrity of various structures. For instance, in plant growth images, the leaves appear to merge or overlap unnaturally, disrupting their biological topology. Similarly, the baselines inaccurately predict parts of the hippocampus, resulting in regions that do not correspond to the original brain structure.

Tab.~\ref{table_comp} reports the evaluation metrics of sample quality and diversity for the generated images from our model IGG and the baselines. While all models achieve similar IS, IGG shows significantly lower FVD, FID, and KID scores, and higher SSIM scores. This indicates that while all models generate diverse images (leading to a good IS), the baselines lack realistic features or proper distribution alignment with real images. The observation of IGG achieving approximately $10$ times better scores demonstrates its ability to generate more realistic images.
\begin{table*}[htbp]
\centering
\caption{A comparison of performance metrics across all methods for various datasets.}
\resizebox{0.9\textwidth}{!}{%
\begin{tabular}{lcccccc}
\toprule
Dataset & Model & FVD $\downarrow$ & FID $\downarrow$ & KID $\downarrow$ & SSIM $\uparrow$ & IS $\uparrow$ \\
\midrule
\multirow{4}{*}{Plant} 
    & IGG & $\textbf{48.29}$ & $\textbf{9.23}$ & $\textbf{0.007} \pm \textbf{0.02}$ & $\textbf{0.98} \pm \textbf{0.02}$ & $\textbf{1.06} \pm \textbf{1.2}\%$ \\
    & VDM~\cite{ho2022video} & $357.56$ & $71.79$ & $0.23 \pm 0.03$ & $0.23 \pm 0.10$ & $1.05 \pm 1.0\%$ \\
    & CogVideoX~\cite{yang2024cogvideox} & $471.49$ & $153.93$ & $0.63 \pm 0.07$ & $0.69 \pm 0.19$ & $1.05 \pm 0.6\%$ \\
    & DynamiCrafter~\cite{xing2025dynamicrafter} & $466.70$ & $80.78$ & $0.41 \pm 0.04$ & $0.89 \pm 0.04$ & $1.03 \pm 0.3\%$ \\
\midrule
\multirow{4}{*}{Brain} 
    & IGG & $\textbf{26.39}$ & $\textbf{6.23}$ & $\textbf{0.006} \pm \textbf{0.008}$ & $\textbf{0.97} \pm \textbf{0.03}$ & $1.02 \pm 0.08\%$ \\
    & VDM~\cite{ho2022video} & $247.84$ & $69.23$ & $0.46 \pm 0.03$ & $0.89 \pm 0.02$ & $1.02 \pm 0.07\%$ \\
    & CogVideoX~\cite{yang2024cogvideox} & $302.20$ & $114.50$ & $0.46 \pm 0.04$ & $0.69 \pm 0.27$ & $\textbf{1.06} \pm \textbf{2.8}\%$ \\
    & DynamiCrafter~\cite{xing2025dynamicrafter} & $288.62$ & $142.40$ & $1.05 \pm 0.03$ & $0.88 \pm 0.03$ & $1.03 \pm 0.20\%$ \\
\bottomrule
\end{tabular}%
}
\label{table_comp}
\end{table*}

\begin{figure*}[!h]
\centering
\includegraphics[width=1.0\textwidth] {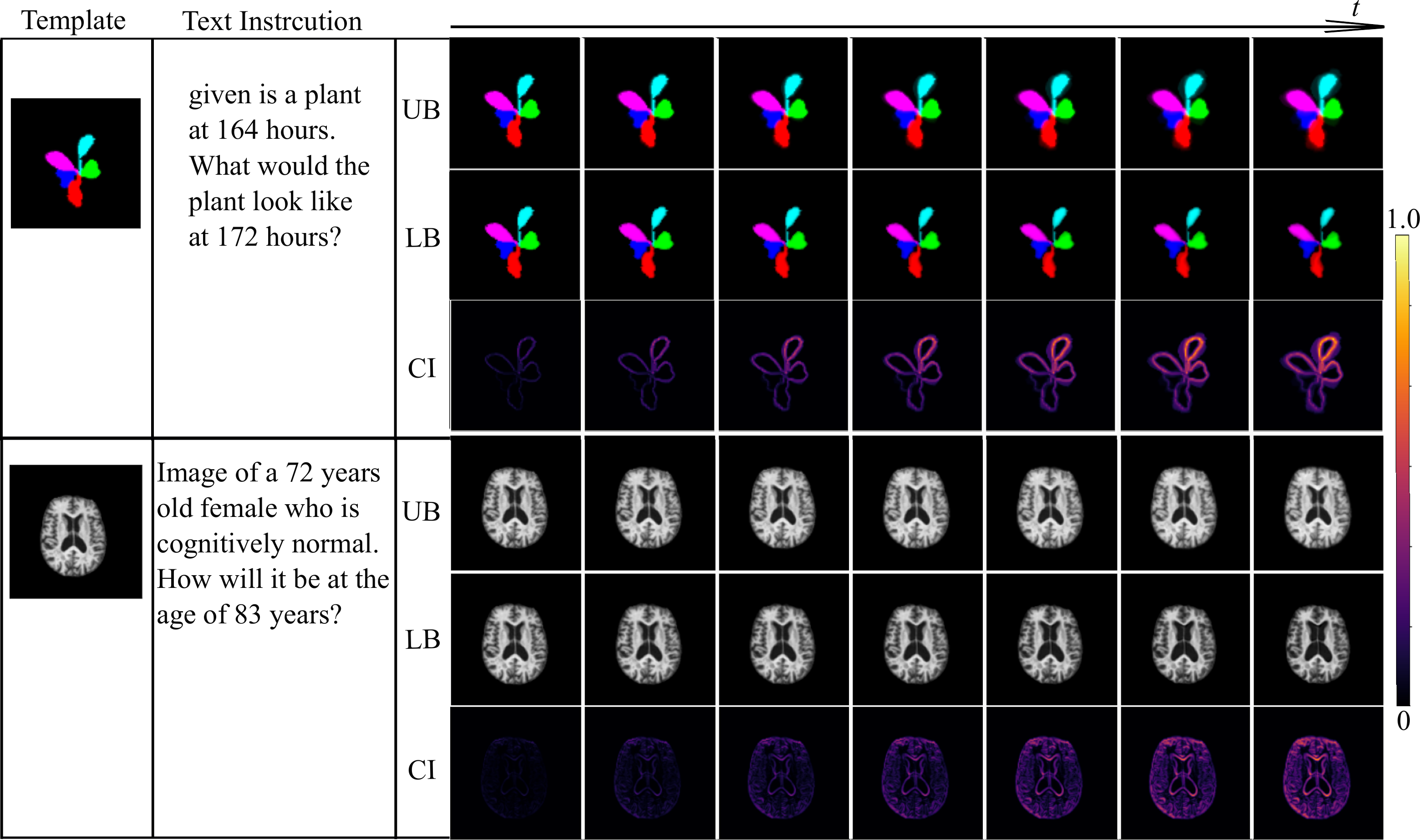}
     \caption{Left to right: input template images with text instructions, followed by confidence maps illustrating the lower bounds (LB), upper bounds (UB), and confidence intervals (CI), which represent regions representing 95\% of ideal growth patterns, based on ~1000 samples generated by our IGG model across different time frames.}
\label{fig:CDM-PANT-OASIS}
\end{figure*}

Fig.~\ref{fig:CDM-PANT-OASIS} visualizes exemplary confidence maps of plant growth and brain progression images generated from IGG. It suggests that our model effectively captures the growth pattern of plants over time, primarily focusing on the boundaries of the leaves. The confidence maps of brain images highlight expanding patterns in ventricles, demonstrating the model's ability to capture dynamic changes over time.

\section{Conclusion \& Discussion}
This paper introduces IGG, a novel approach to image generation that deforms a given template along learned random geodesics in deformation spaces, guided by text instructions. In contrast to current generative models that focus on manipulating image intensity and texture, IGG explicitly learns dynamic geodesics of image transformations during the diffusion process. This enables natural, interpretable transformations and provides quantitative metrics to assess topological changes in generated samples. Our key contributions include a geodesic-informed network to learn latent representations of time-dependent diffeomorphic transformations and a latent geometric diffusion model that captures sequential deformation features conditioned on text inputs. Experimental results demonstrate the effectiveness of IGG with significantly improved sample fidelity and diversity compared to the state-of-the-art. 

Our IGG model advances image generation by integrating geometric principles, offering tools for topology assessment, and establishing a framework for generating anatomically consistent image samples. Future research will further (i) leverage IGG's outputs to benefit downstream tasks, such as image classification, segmentation, and object recognition in the medical domain; and (ii) thoroughly validate and expand its utility across diverse applications.

\paragraph*{\bf Acknowledgments.}
This work was supported by NSF CAREER Grant 2239977.

\bibliographystyle{splncs04}
\bibliography{paper193}

\end{document}